\documentclass[nofootinbib,pra,twocolumn]{revtex4-2}
\usepackage{float,graphicx,multirow, amsmath,color,dsfont}
\usepackage{amsmath,bm}
\usepackage{amssymb,mathrsfs,dsfont}
\usepackage{amssymb,mathbbol}
\usepackage{amsfonts}
\usepackage{mathrsfs}
\usepackage{float} 
\usepackage{xcolor,hyperref}
\definecolor{darkblue}{rgb}{0,0.0.1,0.3}
\definecolor{darkred}{rgb}{0.6,0.1,0}
\hypersetup{colorlinks,breaklinks,
	linkcolor=darkred,urlcolor=darkblue,
	anchorcolor=darkred,citecolor=
	darkred,pdfauthor=Chandan, pdftitle=Chandan.Thermal.Teleportation}
\usepackage{tikz}
\usepackage{mathbbol}
\usepackage{float}
\usepackage[normalem]{ulem}%For bibliography 
\begin{document}

	\title{Non-Gaussian two mode squeezed thermal states in continuous 
		variable quantum teleportation}

	\author{Chandan Kumar}
	\email{chandan.quantum@gmail.com}
	\affiliation{Optics and Quantum Information Group, The Institute of Mathematical Sciences, CIT Campus, Taramani, Chennai 600113, India.}
\affiliation{Homi Bhabha National Institute, Training School Complex, Anushakti Nagar, Mumbai 400085, India.}

\begin{abstract}
While photon catalyzed   two mode squeezed vacuum state has been considered in context of quantum teleportation, similar studies have not been yet conducted  for photon catalyzed two-mode squeezed thermal (TMST) state. This can be attributed to challenges involved in the evaluation of teleportation fidelity for photon catalyzed TMST state. In this article, we consider a practical scheme for the implementation of non-Gaussian operation, viz., photon subtraction, photon addition, and photon catalysis, on TMST state.  The generated states are employed as resources in continuous-variable quantum teleportation. The results show that the  three non-Gaussian operations can enhance the teleportation fidelity. Considering the success probability of the non-Gaussian operations, we identify single-photon catalysis and single photon subtraction to be optimal  for teleporting input coherent states, at low and intermediate squeezing levels.

	\end{abstract}
	\maketitle
	%%%%%%%%%%%%%%%%%%%%%%%%%%%%%%%%%%

	\section{Introduction} 
 
 Squeezed states are an important nonclassical resource for continuous variable (CV) quantum information processing (QIP)~\cite{Braunstein, weedbrook-rmp-2012}.  
Generally pure squeezed states are preferable for quantum tasks;  however, experimental imperfection and   losses degrade the pure state to a mixed state~\cite{tele}. These mixed states form an important family of mixed Gaussian states known as squeezed thermal states. Several studies  characterizing single and multimode squeezed thermal states have been performed. For instance,  photon statistics~\cite{sts1989, statistics,statisics23} and squeezing~\cite{squeezingsts} have been investigated in such states. Reference~\cite{tele} experimentally demonstrated   teleportation of single-mode squeezed thermal state. Entanglement~\cite{enttmst} and tomography~\cite{tmsttomogaphy} of two-mode squeezed thermal (TMST) state has also been undertaken. These states have found practical applications in diverse domains such as quantum teleportation~\cite{thermal-tele}, phase estimation~\cite{thermalPhase, norithermal, chinPB-thermal}, and quantum heat engine~\cite{work}. 

Non-Gaussian operations, namely, photon subtraction (PS), photon addition (PA) and photon catalysis (PC) have been considered on different states to ameliorate performance in various protocols such as quantum metrology~\cite{gerryc-pra-2012,josab-2012,braun-pra-2014,josab-2016,pra-catalysis-2021,crs-ngtmsv-met,manali} and quantum teleportation~\cite{tel2000,tel2009,catalysis15,catalysis17,wang2015,tele-arxiv,better}.
   There exist other types of non-Gaussian operations such as coherent superposition operation 
 	 of photon subtraction and addition~\cite{Hyunchul} and   number-conserving generalized superposition
 	 of products   operation~\cite{PhysRevLett.101.260401,Dhar2015,PhysRevA.103.052602}; however, in this article,   the term non-Gaussian operations shall refer to PA, PS, and PC operations only.

In the context of quantum teleportation, ideal photon-subtracted TMST (PSTMST)~\cite{pstmst,pspatmst} and ideal photon-added TMST (PATMST)~\cite{patmst,pspatmst} states have been considered as resource states. 
However, the usefulness of photon catalyzed TMST (PCTMST) states in quantum teleportation have not been examined. This is largely because of the complicated calculations involved in the 
  evaluation of teleportation fidelity for PCTMST state.

  Further, the ideal PS and PA operations are implemented via annihilation operator  and creation operator, which is non unitary and thus cannot be implemented experimentally. 
   In this article, we consider a practical scheme for the implementation of probabilistic non-Gaussian operations on TMST state [Fig.~\ref{figsub}].   While PS and PA operations generate  PSTMST and  PATMST states, respectively,  PC operation produces photon-catalyzed TMST (PCTMST) states. These three different family of states will be collectively referred as non-Gaussian TMST states. These states are used as resource states in   teleporting input coherent and squeezed vacuum states. 
 The analysis shows that while PS and PC operations can enhance the   fidelity of teleporting input coherent state, PA operation does not. For the teleportation of  input (highly) squeezed vacuum state, all three non-Gaussian operations can enhance fidelity.

 The investigation also reveals that the fidelity via PS operation  is optimized in the unit transmissivity limit, where the PSTMST state is basically reduced to the ideal PSTMST state. In unit transmissivity limit, the success probability of PS operation approaches zero, which is unsuitable from a practical point of view.  Therefore, we trade-off between the enhancement in fidelity and success probability  to find out the optimal non-Gaussian operation. To this end, we consider the product of fidelity enhancement and success probability.   The result shows that  single PC operation and single PS operation on TMST is the optimal operation at low and intermediate squeezing  for the teleportation of input coherent states.

 We stress that our practical scheme for the implementation of non-Gaussian operations can be implemented via currently available technology. 
  The generation of multiphoton Fock state required in the considered practical scheme is feasible~\cite{singlephoton,single-photon,singlephot,2phton,3phton}. Further, photon number resolving detectors can also be    implemented with existing technologies~\cite{Lita:08,Marsili2013,Zadeh}.
 On a theoretical note, the derived Wigner characteristic function of the non-Gaussian TMST state
  is quite general including several parameters such as input multiphoton state, beam splitters transmissivity, and detected number of photons. This renders the calculation of Wigner characteristic function and fidelity a challenging task.

     We arrange the rest of the paper as follow. In sec.~\ref{sec:wig}, we derive the Wigner characteristic function of the non-Gaussian TMST state. In Sec.~\ref{sec:qt}, we explore the advantages of non-Gaussian operations on TMST state in the  teleportation of input coherent and squeezed vacuum states.
In Sec.~\ref{success},  We consider the success probability and identify optimal non-Gaussian operation.
In Sec.~\ref{contour}, we draw two-dimensional plots of success probability and  fidelity enhancement to gain more insights.
  Sec.~\ref{sec:conc} contains a summary of our results and mentions future directions.   Appendix~\ref{intro} contains a short description of CV system and phase space formalism.

	%%%%%%%%%%%%%%%%%%%%%%%%%%%%%%%%%%%%%%%%%%%

	\section{Derivation of the Wigner characteristic function of  non-Gaussian TMST states}\label{sec:wig}

	\begin{figure}[h!]
		\includegraphics[scale=1]{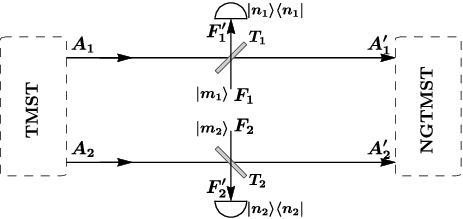}
		\caption{     Schematic for the generation of non-Gaussian two mode squeezed thermal state. Multiphoton states are mixed via beam splitters to each of the mode.  Photon number resolving detectors are employed on both the modes for implementing non-Gaussian operations. }
		\label{figsub}
	\end{figure}

	We  discuss the implementation of non-Gaussian operations on the TMST state. The schematic is shown   in Fig.~\ref{figsub}.
	We start with a TMST state with modes labeled by $A_1$ and $A_2$. A TMST state can be produced by operating two mode squeezing transformation on two uncorrelated thermal modes. The covariance matrix of two uncorrelated thermal modes is given by $V_{\text{th}} = (n_\text{th}+1/2)\mathbb{1}_4$, where $n_\text{th}$  is the average number of photons in a single mode thermal state. Therefore, the covariance matrix of the TMST state can be written as
	\begin{equation}\label{tms}
		V_{A_1A_2}=S_{A_1A_2}(r) V_{\text{th}} S_{A_1A_2}(r)^T, 
	\end{equation}
	where	$S_{A_1A_2}(r)$ is the two mode squeezing transformation given by Eq.~(\ref{tmso}) of the Appendix~\ref{intro}.
	The covariance matrix of the TMSV state can be obtained by setting $n_\text{th}=0$ in Eq.~(\ref{tms}).
	Since TMST state is a Gaussian state with zero mean and covariance matrix given by Eq.~(\ref{tms}),
	the  Wigner characteristic function of the TMST state can be readily evaluated using Eq.~(\ref{wigc}):
	\begin{equation}
		\begin{aligned}
			\chi(\Lambda) =    \exp\big[&- 
			\kappa	(\tau_1^2+\sigma_1^2+\tau_2^2+\sigma_2^2)\cosh (2r)/2   \\
			&+   \kappa (\tau_1 \tau_2-\sigma_1 \sigma_2) \sinh (2r)	\big],
		\end{aligned}
	\end{equation}
	where $ \kappa =(n_\text{th}+1/2)$. 
	Further, modes  $F_1$ and $F_2$ are initialized to Fock states $|m_1\rangle$ and $|m_2\rangle$. The four mode state can be represented via Wigner characteristic function formalism as
	\begin{equation}
		\chi_{F_1 A_1 A_2 F_2}(\Lambda) =  \chi_{A_1 A_2}(\Lambda) \chi_{|m_1\rangle}(\Lambda_3)  \chi_{|m_2\rangle}(\Lambda_4).
	\end{equation}
	Now, mode $A_1$ ($A_2$)  is mixed with mode $F_1$ ($F_2$), using a beam splitter of transmissivity $T_1$ ($T_2$) respectively.  Consequently, the four modes get  entangled. The four-mode entangled state is represented by
	\begin{equation}
		\chi_{F_1' A_1' A_2' F_2'}(\Lambda)  =\chi_{F_1 A_1 A_2 F_2}( B(T_1,T_2)^{-1}\Lambda).
	\end{equation}
	where $ B(T_1,T_2)=B_{A_1 F_1}(T_1) \, {\oplus} \, B_{A_2 F_2} (T_2)$ denotes the collective action of the two beam splitters 
	with $B_{ij}(T)$ being 
	the beam splitter operation given in Eq.~(\ref{bs}) of the 
	Appendix~\ref{intro}. 
	We employ photon number resolving detectors on the auxiliary modes $F_1^{'}$ and $F_2^{'}$. When the detector in the output mode detects $|n_1\rangle$ and $|n_2\rangle$ photons, it heralds successful implementation of non-Gaussian operation on both the modes. The state after the measurement is given by
	\begin{equation}\label{detect}
		\begin{aligned}
			\widetilde{\chi}^{\text{NG}}_{A_1' A_2'}=& \frac{1}{(2 \pi)^2} \int d^2 \Lambda_3  d^2 \Lambda_4 
			\underbrace{\chi_{F_1' A_1' A_2' F_2'}(\Lambda)}_{\text{Four mode entangled state}}\\
			&\times 
			\underbrace{\chi_{|n_1\rangle }(\Lambda_3)}_{\text{Projection on }|n_1\rangle \langle n_1|} 
			\underbrace{\chi_{|n_2\rangle }(\Lambda_4)}_{\text{Projection on }|n_2\rangle \langle n_2|}, \\
		\end{aligned}
	\end{equation}
	which is unnormalized. If $m_i < n_i$, PS operation is performed on mode $A_i$. Similarly,  if $m_i > n_i$ or $m_i = n_i$, PA or PC operation is performed on mode $A_i$.
	Performing PS, PA and PC operations on the TMST state 
	leads to the generation of  PSTMST, PATMST, and PCTMST states, respectively. 
	We can perform non-Gaussian operations either on both modes or on a single mode of the TMST state. The former shall be referred to as symmetric (Sym) non-Gaussian operations, while the latter shall be referred to as asymmetric (Asym) non-Gaussian operations.
	Further, we set $m_1=m_2=0$ for PS operation and  $n_1=n_2=0$ for   PA operation. 
	On integrating Eq.~(\ref{detect}),   we obtain 
	\begin{equation}\label{eqchar}
		\begin{aligned}
			\widetilde{\chi}^{\text{NG}}_{A_1' A_2'}&= \bm{\widehat{F}_1} \exp 
			\left(\bm{\Lambda}^T M_1 \bm{\Lambda}+\bm{u}^T M_2 \bm{\Lambda} + \bm{u}^T M_3 \bm{u} \right),
		\end{aligned}
	\end{equation}
	where the column vectors $\bm{\Lambda}$ and $\bm{u}$ are defined as  
	$(\tau_1,\sigma_1,\tau_2,\sigma_2)^T$ and $(u_1,v_1,u_2,v_2,u_1',v_1',u_2',v_2')^T$ 
	respectively, and the matrices $M_1$, $M_2$ and $M_3$ are provided in 
	Eqs.~(\ref{m1}),~(\ref{m2}) and~(\ref{m3}) of the Appendix~\ref{appdix}. 
	Further,  the differential operator $\bm{\widehat{F}_1} $ is given by 
	\begin{equation}
		\begin{aligned}
			\bm{\widehat{F}_1} = \frac{2^{-(m_1+m_2+n_1+n_2)}}{m_1!m_2!n_1!n_2!} \frac{\partial^{m_1}}{\partial\,u_1^{m_1}} \frac{\partial^{m_1}}{\partial\,v_1^{m_1}} \frac{\partial^{m_2}}{\partial\,u_2^{m_2}} \frac{\partial^{m_2}}{\partial\,v_2^{m_2}}\\
			\times \frac{\partial^{n_1}}{\partial\,u_1'^{n_1}} \frac{\partial^{n_1}}{\partial\,v_1'^{n_1}} \frac{\partial^{n_2}}{\partial\,u_2'^{n_2}} \frac{\partial^{n_2}}{\partial\,v_2'^{n_2}} \{ \bullet \}_{\substack{u_1= v_1=u_2= v_2=0\\ u_1'= v_1'=u_2'= v_2'=0}}.\\
		\end{aligned}
	\end{equation}
	
	Normalization of Eq.~(\ref{eqchar}) yields the success probability of non-Gaussian operations and can be calculated as
	\begin{equation}\label{probng}
		\begin{aligned}
			P^{\text{NG}}&=\widetilde{\chi}^{\text{NG}}_{A_1' A_2'}\bigg|_{\tau_1= \sigma_1= \tau_2= \sigma_2=0}
			= \bm{\widehat{F}_1} \exp 
			\left(\bm{u}^T M_3 \bm{u} \right). \\		
		\end{aligned}
	\end{equation}
	Therefore, the normalized  Wigner characteristic function $\chi^{\text{NG}}_{A'_1 A'_2}$ 
	of the non-Gaussian TMST states turns out to be
	\begin{equation}\label{normPS}
		\chi^{\text{NG}}_{A'_1 A'_2}(\tau_1,\sigma_1,\tau_2,\sigma_2) ={\left(P^{\text{NG}}\right)}^{-1}\widetilde{\chi}^{\text{NG}}_{A_1' A_2'}(\tau_1,\sigma_1,\tau_2,\sigma_2).
	\end{equation}
	We can obtain the Wigner characteristic function of several special states from Eq.~(\ref{normPS}). 
	The Wigner characteristic function of the ideal PSTMST state $ \mathcal{N}_s \hat{a}_1^{n_1}  \hat{a}_2^{n_2} |\text{TMST}\rangle$ is obtained in the unit transmissivity limit $T_1  \rightarrow 1$ and $T_2  \rightarrow 1$   with  $m_1=m_2=0$. Here $\mathcal{N}_a$ is the normalization factor.
	Similarly, the Wigner characteristic function of the ideal PATMST state $ \mathcal{N}_a \hat{a}{_1^{\dagger }}^{m_1} \hat{a}{_2^{\dagger }}^{m_2}  |\text{TMST}\rangle$ is obtained in the unit transmissivity limit $T_1  \rightarrow 1$ and $T_2  \rightarrow 1$   with  $n_1=n_2=0$. Here $\mathcal{N}_s$ is the normalization factor. Further, the Wigner characteristic function of the non-Gaussian TMSV state can be obtained by setting $ \kappa=1/2$ (equivalently $n_\text{th}=0$) in Eq.~(\ref{normPS}).

\section{Advantages of non-Gaussian operations on TMST state in CV quantum teleportation}\label{sec:qt}

We now  analyze the
teleportation of input coherent and squeezed vacuum states using non-Gaussian TMST resource states.
We first briefly describe the BK protocol~\cite{bk-1998} that can be used to teleport an unknown input quantum state between two parties, from Alice to Bob. The implementation of this protocol prerequisites a shared pair of entangled states between the two parties. In order to teleport an unknown single-mode quantum state, Alice uses a balanced beam splitter to combine the input quantum state with her mode of the resource state and subsequently subjects the two output modes to homodyne measurements. The results of these measurements are then communicated by Alice to Bob through a classical channel. As per the results obtained in the measurements, Bob appropriately displaces the mode in his possession in order to retrieve the initial input state. The success of this protocol can be characterized using fidelity $F$, given by,
\begin{equation}
	F =\text{Tr} [\rho_{\text{in}}\rho_{\text{out}}]  
\end{equation}
where $\rho_{\text{in}}$ and $\rho_{\text{out}}$ denote the density operator of the input state and the output state, respectively. In the Wigner characteristic function formalism, where the Wigner characteristic functions of the input and output states are respectively given by $\chi_{\text{in}}(\Lambda_2)$ and $\chi_{\text{out}}(\Lambda_2)$, the fidelity can be computed through the following integral:
\begin{equation}\label{eq:fid}
	F =\frac{1}{2 \pi} \int d^2 \Lambda_2 \chi_{\text{in}}(\Lambda_2) \chi_{\text{out}}(-\Lambda_2),
\end{equation}
where the output state $\chi_{\text{out}}(\Lambda_2)$ $\equiv$ $\chi_{\text{out}}(\tau_2,\sigma_2)$, can be expressed as a product function of the Wigner characteristic function of the input state and the entangled resource state:

\begin{equation}
	\chi_{\text{out}}(\tau_2,\sigma_2) = \chi_{\text{in}}(\tau_2,\sigma_2) \chi_{A_1' A_2'}(\tau_2,-\sigma_2,\tau_2,\sigma_2),
\end{equation}

It is a known result that the fidelity of teleporting a coherent state cannot exceed the value of 1/2 when only classical resources are utilized~\cite{Braunstein-jmo-2000,Braunstein-pra-2001}. Therefore, a fidelity value above 1/2 indicates the usage of quantum resources.

\subsection{Teleporting an input coherent state using non-Gaussian TMST resource states}

Having described the BK protocol for CV quantum teleportation, we now consider the teleportation of input coherent state. To evaluate the fidelity, we use the Wigner characteristic function of the non-Gaussian TMST resource states~(\ref{normPS}) and coherent state~(\ref{wigc}). The expression of fidelity can be evaluated  using Eq.~(\ref{eq:fid}). We proceed to analyze the fidelity of teleporting input coherent state using   non-Gaussian TMST resource states with respect to squeezing and thermal parameters.

\begin{figure}[htbp]
	\begin{center}
		\includegraphics[scale=1]{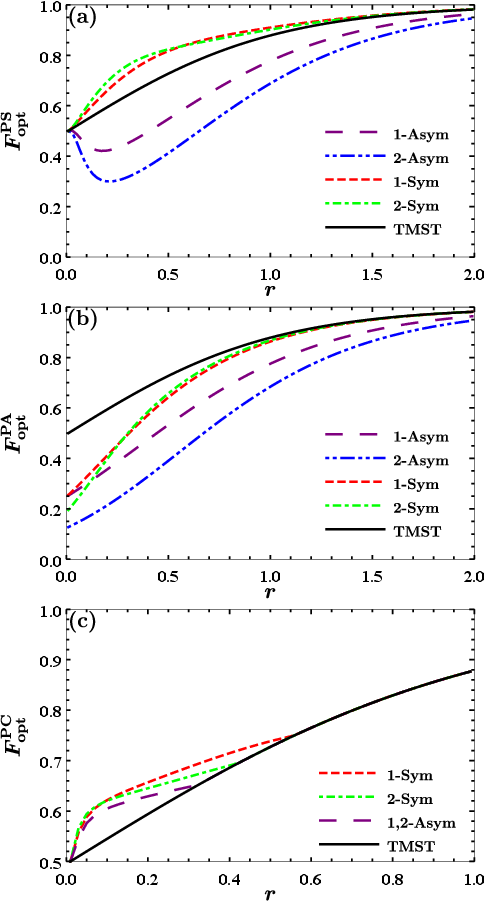}
		\caption{Optimized fidelity for teleporting input coherent state
			as a function of the squeezing parameter $r$ for (a) PSTMST
			states, (b) PATMST states, and (c) PCTMST states with $\kappa=0.51$. We have adjusted the transmissivities of the beam splitters   to maximize the fidelity.
		}
		\label{fidc}
	\end{center}
\end{figure}

While annalyzing the fidelity as a function of the squeezing parameter, we consider a fixed low value of thermal parameter $\kappa = 0.51$ to make the fidelity curves for different non-Gaussian TMST states distinguishable. Different fidelity curves coincide for higher $\kappa$ values and therefore are difficult to  distinguish\footnote{We  explore the fidelity dependence on $\kappa$ in Sec.~\ref{sec:thermal}.}.

We perform numerical optimization to maximize the fidelity by adjusting the transmissivity of the beam splitters. The results of this optimization are presented in Fig.~\ref{fidc}. We observe that Sym PS operations enhance the teleportation fidelity of the TMST state. However, Asym PS operations are detrimental to quantum teleportation.  We also see that neither Asym nor Sym PA operation enhances the teleportation fidelity.  The analysis of Sym PC operation reveals that it can enhance the fidelity till a certain squeezing threshold $r_{\text{th}}$. Beyond this threshold squeezing, the fidelity is optimized at unit transmissivity. Since the PCTMST state becomes the TMST state in unit transmissivity limit, the fidelity beyond  $r_{\text{th}}$ for the PCTMST state is equal to that of the TMST state. While performing PC operation on one mode is undesirable, we also see that Asym 1,2-PC operation ($m_1=n_1=1$ and $m_2=n_2=2$) is beneficial. 

We observe a major difference between the fidelity results for PC operation on  TMST and  TMSV~\cite{tele-arxiv} states. While the optimized fidelity for Sym 1-PCTMSV  state exhibits a jump as the squeezing transitions from zero to non-zero values,  no such characteristic is observed for Sym 1-PCTMST state.
\begin{figure}[htbp]
	\begin{center}
		\includegraphics[scale=1]{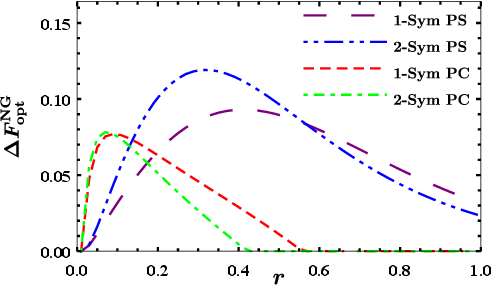}
		\caption{Optimized fidelity enhancement,    $\Delta F^{\text{NG}}$, as a function of the squeezing parameter $r$  with $\kappa=0.51$. We have adjusted the transmissivities of the beam splitters   to maximize  
			$\Delta F^{\text{NG}}$.}
		\label{dfidc}
	\end{center}
\end{figure}

To get a precise idea of the magnitude of enhancement and the optimal  squeezing and transmissivity, we can consider a new figure of merit termed fidelity enhancement   defined as follows:
\begin{equation}
	\Delta F^{\text{NG}}  = F^{\text{NG}}- F^{\text{TMST}}.
\end{equation}
We optimize $\Delta F^{\text{NG}}$ with respect to transmissivity  and present the results  in Fig.~\ref{dfidc}. We see that the enhancement rendered by Sym PS operations surpasses the Sym PC operations. 
Further, the 2-Sym PS operation demonstrates superior performance when compared to the 1-Sym PS operation. On the other hand, both the 1-Sym and 2-Sym PC operations exhibit nearly identical performance.

\begin{figure}[htbp]
	\begin{center}
		\includegraphics[scale=1]{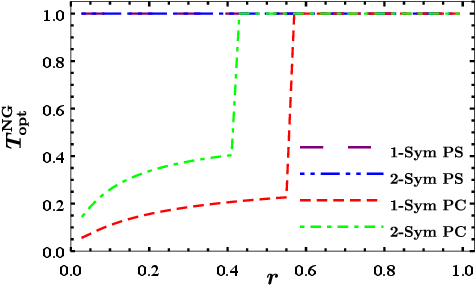}
		\caption{The optimal beam splitter transmissivities, that maximize $F^{\text{NG}}$ (Fig.~\ref{fidc})  or $\Delta F^{\text{NG}}$ (Fig.~\ref{dfidc}).}
		\label{dfidct}
	\end{center}
\end{figure}

The optimal transmissivity values for Figs.~\ref{fidc} and \ref{dfidc} are identical and are depicted in Fig.~\ref{dfidct}.
The results reveal that the optimal transmissivity for PS operation is unity, and therefore the corresponding fidelity results are identical to those of the ideal PSTMST states. Similarly, the fidelity for the PATMST state is optimized in unit transmissivity limit ( not shown in Fig.~\ref{dfidct}) and hence the results are identical to those of the ideal PATMST state.  For the PC operation, the optimal transmissivity lies below 1/2 until a certain threshold squeezing value, denoted as $r_{\text{th}}$, beyond which the optimal transmissivity becomes unity.
It is worth mentioning that  previous works, such as Refs.~\cite{pstmst,patmst,pspatmst} have already investigated quantum teleportation employing ideal PSTMST and PATMST resource states.

\subsection{Teleporting an input squeezed vacuum state using non-Gaussian TMST resource states}
\begin{figure}[htbp]
	\begin{center}
		\includegraphics[scale=1]{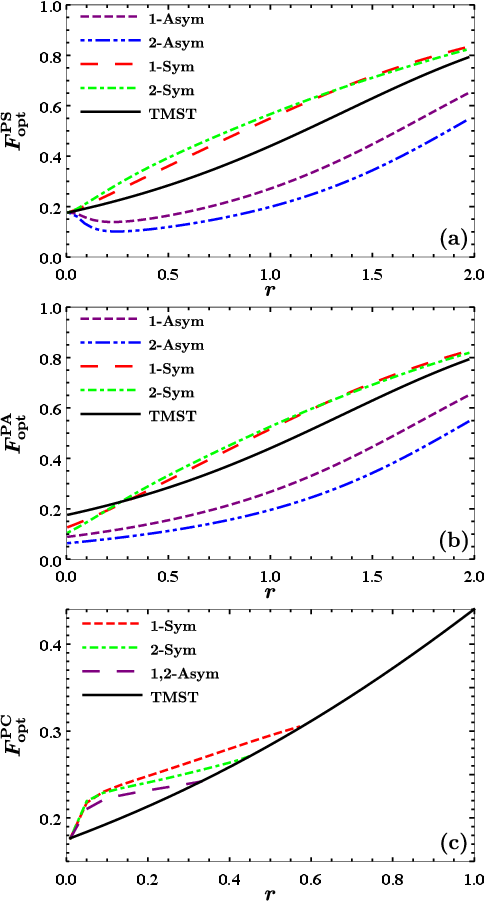}
		\caption{Optimized fidelity   as a function of the squeezing parameter $r$ for (a) PSTMST states, (b) PATMST states, and (c) PCTMST states for $\kappa=0.51$. The state to be teleported is input squeezed vacuum state with $\epsilon=1.7$. We have adjusted the transmissivities of the beam splitters   to maximize the fidelity. 
		}
		\label{fids}
	\end{center}
\end{figure}
We now consider the teleportation of an input squeezed vacuum state with 
squeezing $\epsilon$, whose Wigner characteristic function is provided in Eq.~(\ref{chi_sqv}).  The expression of  fidelity can be evaluated using the expression of fidelity~(\ref{eq:fid}).
For our analysis, we choose the squeezing of the input vacuum state   to be $\epsilon=1.7$, which  is the maximum squeezing  achieved experimentally up until now~\cite{15dB}.
We maximize the fidelity by adjusting the transmissivity of the beam splitters and present the results in Fig.~\ref{fids}. The fidelity curves corresponding to PS and PC operations display similar behavior to that was observed for the teleportation of an input coherent state (Fig.~\ref{fidc}). However, we observe that Sym PA operations on the TMST state can also enhance the fidelity. This result is contrary to the earlier observation of teleportation of an input coherent state.

\begin{figure}[htbp]
	\begin{center}
		\includegraphics[scale=1]{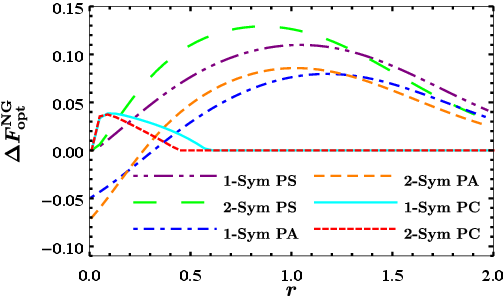}
		\caption{ Optimized fidelity enhancement   $\Delta F^{\text{NG}} $, as a function of the squeezing parameter $r$  with $\kappa=0.51$.  The state to be teleported is input squeezed vacuum state with $\epsilon=1.7$. We have adjusted the transmissivities  to maximize $\Delta F^{\text{NG}}$.
		}
		\label{dfids}
	\end{center}
\end{figure}
To identify the optimal values of squeezing and transmissivity that yield maximum advantage, we employ fidelity enhancement as a figure of merit. The results are shown in Fig.~\ref{dfids}. We observe that PS operations offer the greatest advantage,  followed by PA operations. However, at low squeezing, PC operations provide the highest advantage.

\begin{figure}[htbp]
	\begin{center}
		\includegraphics[scale=1]{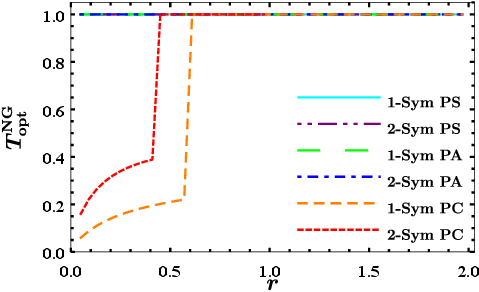}
		\caption{  The optimal beam splitter transmissivities, that maximize $F^{\text{NG}}$ (Fig.~\ref{fids})  or $\Delta F^{\text{NG}}$ (Fig.~\ref{dfids}). }
		\label{dfidst}
	\end{center}
\end{figure}

We  plot the transmissivity maximizing the fidelity (or fidelity enhancement) in Fig.~\ref{dfidst}.
The trend observed in the optimal transmissivity closely aligns with that of coherent state teleportation (Fig.~\ref{dfids}).
The optimal transmissivity for the PS and PA operation turns out to be unity, resulting in what we term as ideal PSTMST and ideal PATMAST states. These states have been employed as resource states for teleportation of input squeezed vacuum state in Ref.~\cite{pspatmst}.

\subsection{Success probability and optimal non-Gaussian operation  }\label{success}

As stated in Sec.~\ref{sec:wig}, the considered non-Gaussian operations exhibit a probabilistic nature, with their probability being determined by Eq.~(\ref{probng}). In our previous analysis, our primary focus was to maximize the fidelity by adjusting the transmissivity  without considering success probability.
To   emphasize the significance of considering success probability, let us examine the case of 1-PSTMST state.  It was observed that the fidelity is maximized in the unit transmissivity limit. However, in this particular limit, the success probability approaches zero. This scenario is highly undesirable from a practical point of view.

To obtain optimal conditions, we trade off between the enhancement in fidelity $\Delta F^{\text{NG}}$ and  success probability by adjusting the transmissivity.  To perform a thorough quantitative analysis, we examine the product $\mathcal{R}^{\text{NG}}=   \Delta F^{\text{NG}}  \times   P^{\text{NG}}$. We aim to  find its maximum value and the corresponding values of squeezing and transmissivity. We have shown the results in Table~\ref{table1} for the teleportation of input coherent states.  
 On comparing with Fig.~\ref{dfidc},   the magnitude of $\Delta F^{\text{NG}}$ has reduced from $9.3 \times 10^{-2}$ to $3.3 \times 10^{-2}$ for the 1-PSTMST state. However,     the success probability, which tends to zero in the unit transmissivity limit,  is now of the order $  10^{-2}$ -- a practically reasonable value.  It is worth mentioning that an experimental   demonstration of single PS operation with an even lower  probability (of the order $  10^{-5}$)   has been conducted~\cite{subprob}.

\begin{table}[htbp]
	\centering
	\caption{\label{table1}
		The maximum value of the product 	$\mathcal{R}^{\text{NG}}=   \Delta F^{\text{NG}}  \times   P^{\text{NG}}$  and magnitude of the other quantities at optimal squeezing and transmissivity. }
	\renewcommand{\arraystretch}{1.5}
	\begin{tabular}{ |c |c |c|c|c |}
		%	\hline \hline
		\hline \hline
		States &  1-PSTMST & 1-PSTMSV &  1-PCTMST &1-PCTMSV \\   \hline \hline
		$\mathcal{R}^{\text{NG}}_{\text{max}} $ & 8.2 $\times 10^{-4}$ & 9.5 $\times 10^{-4}$ & 2.2  $\times 10^{-3}$ & 2.9 $\times 10^{-3}$ \\
		$r_{\text{opt}}$ &0.64 & 0.64&0.24 & 0.26 \\
		$T_{\text{opt}}$ & 0.78&0.77 &0.18 & 0.18 \\
		
		$F^{\text{NG}} $ & 0.81& 0.82 &0.66 &0.70 \\
		$\Delta F^{\text{NG}} $ &3.3 $\times 10^{-2}$ & 3.7 $\times 10^{-2}$& 5.5 $\times 10^{-2}$& 7.1 $\times 10^{-2}$\\
		$P^{\text{NG}} $ &2.5 $\times 10^{-2}$ & 2.6 $\times 10^{-2}$&4.0 $\times 10^{-2}$ & 4.1 $\times 10^{-2}$ \\
		
		\hline \hline
	\end{tabular}
\end{table}

\begin{figure}[htbp]
	\begin{center}
		\includegraphics[scale=1]{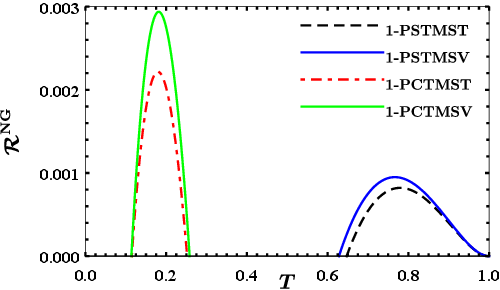}
		\caption{   Product 
			$\mathcal{R}^{\text{NG}}=  P^{\text{NG}} \times  \Delta F^{\text{NG}} $  as a function of the transmissivity $T$ for different non-Gaussian TMST and non-Gaussian TMSV states for teleportation of input coherent states. We have taken $\kappa=0.51$ for non-Gaussian TMST states and the optimal squeezing value is considered (Table~\ref{table1}). Further, we have considered symmetric non-Gaussian operations.}
		\label{quant}
	\end{center}
\end{figure}

For a visual perspective, we   plot the product $\mathcal{R}^{\text{NG}}$ as a function of transmissivity for different non-Gaussian TMSV and non-Gaussian TMST states at optimal squeezing in Fig~\ref{quant}. We observe that while the PC operation is optimal at low squeezing, the PS operation is optimal at intermediate squeezing.  We can verify the   numerical values of different quantities in Fig.~\ref{quant} with numerical values given in Table~\ref{table1}.

 In the next section, we delve into an exploration of how the fidelity varies with the thermal parameter.

\section{Variation of fidelity with thermal parameter $\kappa$}\label{sec:thermal}

\begin{figure}[h!]
	\begin{center}
		\includegraphics[scale=1]{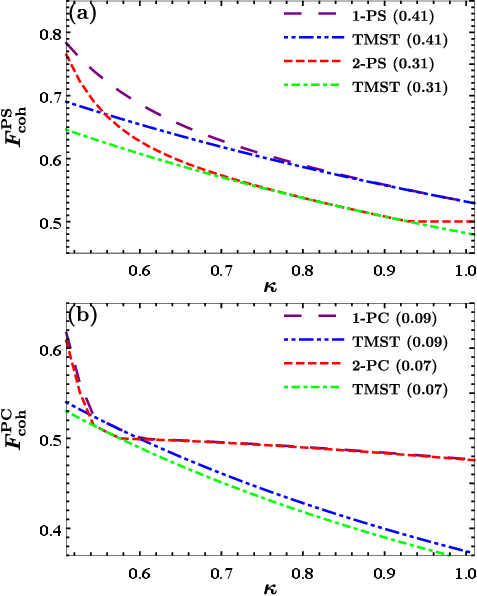}
		\caption{   Optimized fidelity for teleporting input coherent state
			as a function of the thermal parameter $\kappa$ for (a) PSTMST
			states,   and (b) PCTMST states. The optimal squeezing $r$  has been set in accordance with  Fig.~\ref{dfidc}. We have adjusted the transmissivities of the beam splitters   to maximize the fidelity. We note that we have considered symmetric non-Gaussian operations.}
		\label{cfidk}
	\end{center}
\end{figure}

In the earlier analysis, we have set the thermal parameter $\kappa = 0.51$. In this section, we study the variation of the fidelity with $\kappa$.  We plot the optimized fidelity for teleporting input coherent states as a function of $\kappa$ for PSTMST and PCTMST states in Fig.~\ref{cfidk}. We observe that both Sym 1-PS and Sym 2-PS operations can enhance the fidelity in the initial range of the thermal parameter. Moreover, we also notice that Sym 2-PS operation can enhance the fidelity for $\kappa \gtrapprox 0.95$. The corresponding optimal transmissivity  for 2-PS operation turns out to be zero.   In this region, the numerical value of  the fidelity is equal to 1/2.

In the case of PC operation, the fidelity for PCTMST states surpasses that of the TMST state in the initial thermal parameter range. This is followed by a narrow thermal parameter region where the fidelity of the PCTMST states matches that of the TMST state.  In this narrow region, the optimal transmissivity  is unity, where the PCTMST states are basically the TMST state. As the thermal parameter increases, the PCTMST states consistently outperform the TMST state in terms of fidelity.   However, within this particular region, the fidelity falls below the threshold of  1/2, thereby demonstrating that quantum resources are not useful.

Nevertheless, within this particular domain, it is worth noting that the fidelity falls below the threshold of 1/2, thereby exemplifying the limited utility of quantum resources in this context.

\begin{figure}[h!]
	\begin{center}
		\includegraphics[scale=1]{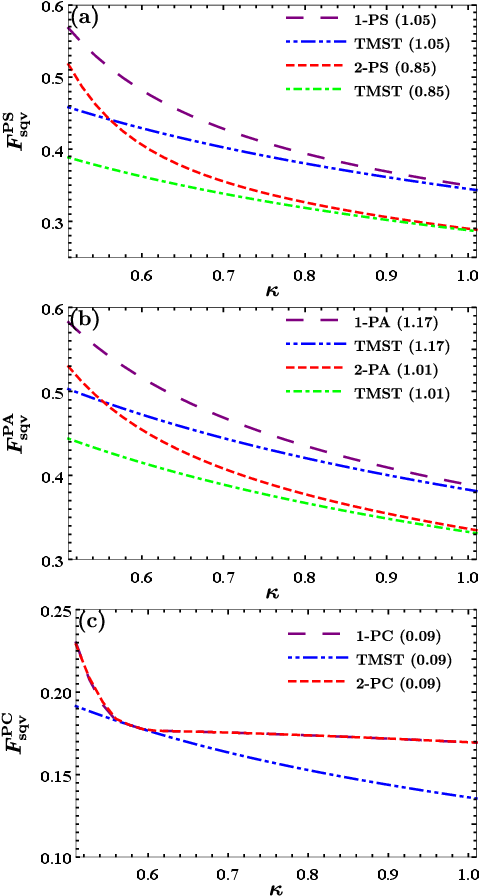}
		\caption{Optimized fidelity for teleporting input squeezed vacuum state
			as a function of the thermal parameter $\kappa$ for (a) PSTMST
			states, (b) PATMST states,  and (c) PCTMST states. The optimal squeezing $r$  has been set in accordance with  Fig.~\ref{dfids}. We have set the squeezing of the input squeezed vacuum state $\epsilon=1.7$. We have adjusted the transmissivities of the beam splitters   to maximize the fidelity. The value of squeezing is provided in parentheses. We note that we have considered symmetric non-Gaussian operations. }
		\label{sfidk}
	\end{center}
\end{figure}

\begin{figure}[h!]
	\begin{center}
		\includegraphics[scale=1]{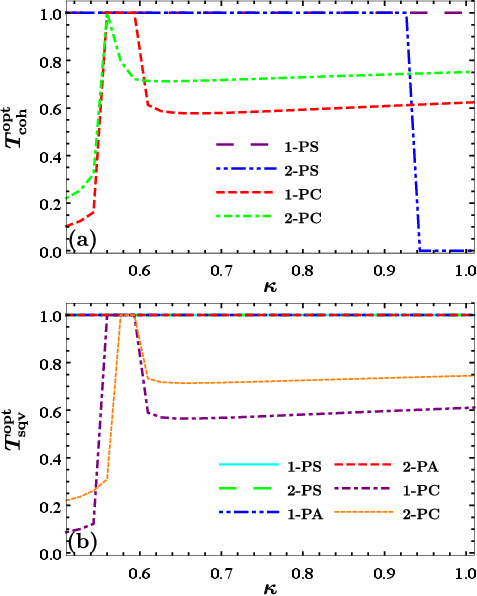}
		\caption{Optimal transmissivity  as a function of the thermal parameter $\kappa$ for (a) teleportation of input coherent states, and (b) teleportation of input squeezed vacuum states.  } 
		\label{tfidk}
	\end{center}
\end{figure}

Now let us focus on the scenario where we teleport   input squeezed vacuum states and conduct a similar analysis.
The results of this analysis are shown in Fig.~\ref{sfidk}. The   fidelity trends for both PS and PC operations are more or less similar to those observed in the case of input coherent state teleportation. However,   the PS   operation  can enhance the fidelity for a larger range of thermal parameter range as compared to the case of input coherent state teleportation. Additionally, we note that PA operation can enhance the fidelity within the considered range of the thermal parameter. As depicted in Fig.~\ref{tfidk}(b), the optimal transmissivity associated with PS and PC operations in the context of teleporting input squeezed vacuum states  follows a similar pattern to that observed in the teleportation of input coherent states (Fig.~\ref{tfidk}(a)). Furthermore, the optimal transmissivity for PA operations is unity.

	\section{Two dimensional  plot of success probability and fidelity enhancement }\label{contour}
\begin{figure}[h!]
	\begin{center}
		\includegraphics[scale=1]{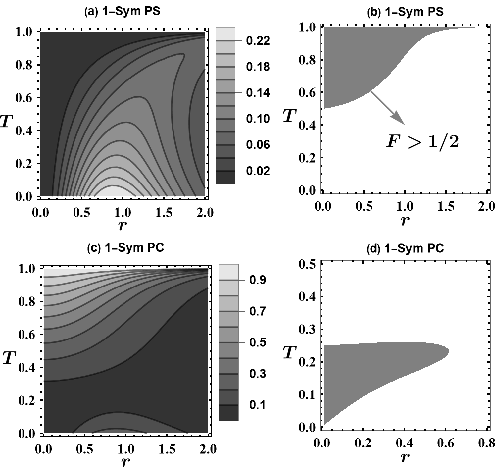}
		\caption{First column: Success  probability  of different non-Gaussian operations on the TMSV state. Second column: Fidelity enhancement $\Delta F^{\text{NG}}$ for the teleportation of input coherent states using different non-Gaussian TMSV resource states. Gray shaded region corresponds to  $\Delta F^{\text{NG}}>0$ with $F>1/2$. }
		\label{ngtmsv}
	\end{center}
\end{figure}
We now analyze the advantageous  non-Gaussian resource states namely 1-PSTMST and 1-PCTMST states for teleporting input coherent state using two-dimensional plots.    This approach provides us insightful observations about the magnitude of success probability and  fidelity enhancement $\Delta F^{\text{NG}}$ for different squeezing and transmissivity values.

To see the effects of thermal parameter, we first carry out the analysis for NGMTSV states (special case of non-Gaussian TMST states obtained for $\kappa=1/2$) and present the results in Fig.~\ref{ngtmsv}.

 In the first column, we depict the success probability of different non-Gaussian operations. We note that the maximum success probability attainable for PS operation is less than PC operation.  In the second column,  the fidelity enhancement  $\Delta F^{\text{NG}}$ for teleporting input coherent   state is shown.
 Gray shaded region corresponds to  $\Delta F^{\text{NG}}>0$ with $F>1/2$. For squeezing and transmissivity in the gray shaded parameter region, it is advantageous to  perform non-Gaussian operations.   A detailed quantitative analysis for non-Gaussian TMSV states can be found in Ref.~\cite{tele-arxiv}.

\begin{figure}[t]
	\begin{center}
		\includegraphics[scale=1]{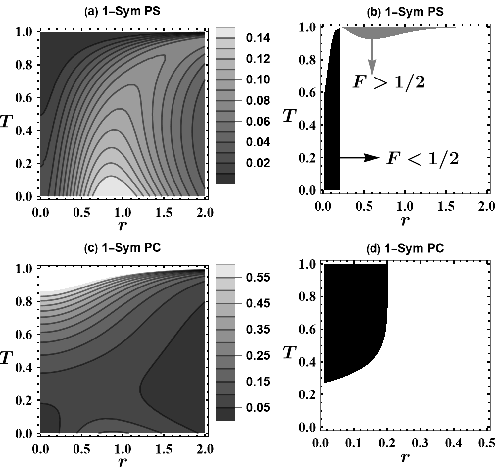}
		\caption{First column: Success  probability  of different non-Gaussian operations on the TMST state. Second column: Fidelity enhancement $\Delta F^{\text{NG}}$ for the teleportation of input coherent states using different non-Gaussian TMST resource states. Gray (black) shaded region corresponds to  $\Delta F^{\text{NG}}>0$  with $F>1/2$ ($F<1/2$). We have taken $\kappa=0.75$.}
		\label{ngtmstmid}
	\end{center}
\end{figure}

\begin{figure}[t]
	\begin{center}
		\includegraphics[scale=1]{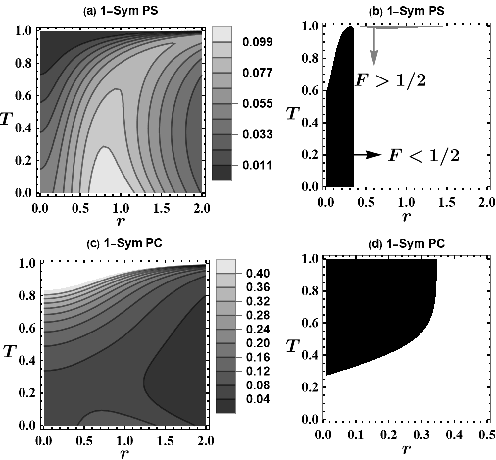}
		\caption{First column: Success  probability  of different non-Gaussian operations on the TMST state. Second column: Fidelity enhancement $\Delta F^{\text{NG}}$ for the teleportation of input coherent states using different non-Gaussian TMST resource states. Gray (black) shaded region corresponds to  $\Delta F^{\text{NG}}>0$  with $F>1/2$ ($F<1/2$).  We have taken $\kappa=1$.}
		\label{ngtmst}
	\end{center}
\end{figure}
We shall now embark upon the examination of the success probaility and fidelity enhancement rendered by the non-Gaussian TMST states. We present the results in Figs.~\ref{ngtmstmid} and~\ref{ngtmst} for $\kappa =3/4$ and 1, respectively. We note that the magnitude of success probability of non-Gaussian operations on TMST states is reduced as compared to non-Gaussian operations on TMSV states.

We now discuss the fidelity enhancement $\Delta F^{\text{NG}}$ for the teleportation of input coherent state. In addition to the gray shaded region, we see the appearance of black  shaded region corresponds to  $\Delta F^{\text{NG}}>0$  but   $F<1/2$.
We see that a vertical straight line along the transmissivity ordinate divides the  black shaded region to the other parameter region.  

 Here, we offer an explanation for the emergence of  straight lines and $F<1/2$ in the case of the PSTMST state (Figs.~\ref{ngtmstmid}(b) and ~\ref{ngtmst}(b)). The classical nature of the TMST state persists until the condition $e^{-2r} \kappa < 1/2$ is satisfied. Specifically, for $\kappa = 3/4$ and 1, the TMST state is a classical state when $r$ is less than 0.20 and 0.35, respectively. Since a PS operation is incapable of  transforming a classical state into nonclassical state~\cite{Zelaya}, the teleportation fidelity via classical resource state remains lower than 1/2.

 \section{Conclusion}\label{sec:conc}
 In this study, we explored the significance of PS, PA, and PC operations on the TMST state  in the context of quantum teleportation. Our findings demonstrate that these non-Gaussian operations can   improve the performance of quantum teleportation. Additionally, our practical scheme allows us to factor in the success probability. When accounting for success probability, we observe that single PC and single PS operations prove to be optimal at low and intermediate squeezing levels for the quantum teleportation of input coherent states.

   Our analysis addresses questions about whether the PC operation can enhance teleportation fidelity and how it compares with PS and PA operations. This analysis also helped answer the problem posed in posed in Ref.~\cite{pspatmst} regarding the utility of the PCTMST state in quantum teleportation.

 While TMSV states are more advantageous for quantum tasks, such states may not exist as  pure states degrade to mixed states due to experimental imperfections and inevitable losses. The transformed mixed  state is similar to the TMST state, rendering the analysis of TMST states important~\cite{tele,thermal-2009,thermal-prl-14,adesso-thermal-17,lossythermal,thermal-metro}. 
 
The Wigner characteristic function for the non-Gaussian TMST states derived in this work is not available in the existing literature to the best of our knowledge. The introduction of this particular expression will serve as a valuable addition to the literature and prove indispensable in the exploration of CV QIP protocols dealing with non-Gaussian TMST states. It will also be useful in the characterization of non-Gaussian TMST via quantifying nonlocality~\cite{nonlocality}, steering~\cite{steering}, entanglement~\cite{ent-prl-2010}, non-Gaussianity~\cite{Ivan2012,ng-2021} and nonclassicality~\cite{nc-2020}.

	\appendix

	\section{Phase space approach to CV systems}\label{intro}

	Our system of concern is a multi-mode system consisting of $n$ non-interacting bosonic modes. The combined Hilbert space of such a system is given by the tensor product of the Hilbert spaces corresponding to the individual modes, $\mathcal{H}^{\otimes n} = \otimes_{i=1}^{n}\mathcal{H}_i$ 
	Here, $\mathcal{H}_i$ denotes the Hilbert space of the  $i^{\text{th}}$ mode.
	We represent the quadrature operators corresponding to the ith mode of our system by $\hat{q}_i$ and $\hat{p}_i$ . Then the set of quadrature operators of our  $n$-mode system can be written as~\cite{arvind1995,Braunstein,adesso-2007,weedbrook-rmp-2012,adesso-2014}
	\begin{equation}\label{eq:columreal}
		\hat{ \xi}  =(\hat{ \xi}_i)= (\hat{q_{1}},\,
		\hat{p_{1}}, \dots, \hat{q_{n}}, 
		\, \hat{p_{n}})^{T}, \quad i = 1,2, \dots ,2n.
	\end{equation}
	The commutation relation between these quadrature operators can be expressed as (in natural units)
	\begin{equation}\label{eq:ccr}
		[\hat{\xi}_i, \hat{\xi}_j] = i \Omega_{ij}, \quad (i,j=1,2,...,2n),
	\end{equation}
	where  
	\begin{equation}
		\Omega = \bigoplus_{k=1}^{n}\begin{pmatrix}
			0& 1\\
			-1&0 
		\end{pmatrix}  
	\end{equation}
	is the symplectic form on $n$ modes.
	
	Further, we define the photon annihilation operator for the ith mode as, $\hat{a}_i=   \frac{1}{\sqrt{2}}(\hat{q}_i+i\hat{p}_i)$. In this article, two Gaussian operations namely, beam splitter operation $B_{ij}(T)$ and two mode squeezing operation $S_{ij}(r)$ is of special interest. The action of these Gaussian  operations on the quadrature operators can be described through the following transformation matrices:
	
	\begin{equation}\label{bs}
		\begin{aligned}
			B_{ij}(T):\left[\begin{array}{c}
				\hat{\xi}_i\\
				\hat{\xi}_j
			\end{array}\right] \mapsto\left[\begin{array}{cc}
				\sqrt{T} \,\mathbb{1}_2 & \sqrt{1-T} \,\mathbb{1}_2\\
				-\sqrt{1-T} \,\mathbb{1}_2 & \sqrt{T} \,\mathbb{1}_2
			\end{array}\right]\left[\begin{array}{l}
				\hat{\xi}_i\\
				\hat{\xi}_j
			\end{array}\right], \\
		\end{aligned}
	\end{equation}
	\begin{equation}\label{tmso}
		\begin{aligned}
			S_{ij}(r):\left[\begin{array}{c}
				\hat{\xi}_i\\
				\hat{\xi}_j
			\end{array}\right] \mapsto\left[\begin{array}{cc}
				\cosh r \,\mathbb{1}_2 & \sinh r \,\mathbb{Z} \\
				\sinh r \,\mathbb{Z} &  \cosh r \,\mathbb{1}_2
			\end{array}\right]\left[\begin{array}{l}
				\hat{\xi}_i\\
				\hat{\xi}_j
			\end{array}\right],
		\end{aligned}
	\end{equation}
	where $\hat{\xi} = (\hat{q}_i,\hat{p}_i)^T$, $\mathbb{1}_2$ is the $2 \times 2$ identity matrix and $\mathbb{Z} = \text{diag}(1,\, -1)$.
	The TMST state is generated by applying the two mode squeezing operator on two uncorrelated thermal modes.

	Now we introduce the 
	Wigner characteristic function for representing a given density operator in the phase space formalism. 
	The Wigner 
	characteristic function of an $n$-mode CV system with density operator $\hat{\rho}$ 
	is given by
	\begin{equation}\label{wigdef}
		\chi(\Lambda) = \text{Tr}[\hat{\rho} \, \exp(-i \Lambda^T \Omega \hat{\xi})],
	\end{equation}
	where $\xi = (\hat{q_1}, \hat{p_1},\dots \hat{q_n}, \hat{p_n})^T$,  
	$\Lambda = (\Lambda_1, \Lambda_2, \dots \Lambda_n)^T$ with  
	$\Lambda_i = (\tau_i, \sigma_i)^T \in \mathcal{R}^2$.

	It is worth noting that both the density operator formalism and the phase space approach using Wigner characteristic function are equivalent descriptions of CV systems. However in this article, we have exclusively used the latter for the sake of mathematical convenience. 
	
	For a single mode Fock state $|n\rangle$, the corresponding Wigner characteristic function can be obtained through Eq.~(\ref{wigdef}) and turns out to be:
	\begin{equation}\label{charfock}
		\chi_{|n\rangle}(\tau,\sigma)=\exp  \left[- \frac{\tau^2}{4}-\frac{\sigma^2}{4} \right]\,L_{n}\left( \frac{\tau^2}{2}+\frac{\sigma^2}{2} \right).
	\end{equation}
	In the above equation, $L_n(x)$ is the Laguerre polynomial which belongs to the set of the classical orthogonal polynomials. The Laguerre polynomial in the above equation can be substituted with its corresponding generating function:
	\begin{equation}\label{charfock1}
		\chi_{|n\rangle}(\tau,\sigma)=\exp  \left[- \frac{\tau^2}{4}-\frac{\sigma^2}{4} \right]\,	\bm{\widehat{F}}e^{ 2 st +s(\tau+i\sigma)-t(\tau-i\sigma)},
	\end{equation}
	with
	\begin{equation}
		\bm{\widehat{F}} =  \frac{1}{2^n n!}  \frac{\partial^n}{\partial\,s^n} \frac{\partial^n}{\partial\,t^n} \{ \bullet \}_{s=t=0}.
	\end{equation}
	The first order quadrature moments associated with our $n$-mode CV system described by the density operator $\hat{\rho}$ can be defined as
	\begin{equation}\label{eq:dis}
		\bm{d} = \langle  \hat{\xi } \rangle =
		\text{Tr}[\hat{\rho} \hat{\xi}].
	\end{equation}
	The second order quadrature moments can be represented in the form of the following matrix:
	\begin{equation}\label{eq:cov}
		V = (V_{ij})=\frac{1}{2}\langle \{\Delta \hat{\xi}_i,\Delta
		\hat{\xi}_j\} \rangle,
	\end{equation}
	where $\Delta \hat{\xi}_i = \hat{\xi}_i-\langle \hat{\xi}_i
	\rangle$, and $\{\,, \, \}$ denotes anti-commutator.
	The above matrix, termed as covariance matrix, is a symmetric square matrix of order $2n$ with all of its entries real.

	The states for which the quasi-probability distributions take on a Gaussian form are termed as Gaussian states. As a result, the characteristic functions corresponding to these distributions also turn out to be Gaussian as they are related to their corresponding distributions via a Fourier transform. These states are relatively simpler for analysis as their first and second order moments are sufficient to describe them. As stated earlier, in this article, we have utilized the Wigner characteristic function approach as it eases the mathematical complexity. We can write the Wigner characteristic function for a Gaussian state as follows~\cite{weedbrook-rmp-2012, olivares-2012}:
	\begin{equation}\label{wigc}
		\chi(\Lambda) =\exp[-\frac{1}{2}\Lambda^T (\Omega V \Omega^T) \Lambda- i (\Omega \bm{d} )^T\Lambda].
	\end{equation}
	Here $\bm{d}$ and $V$ have their usual meanings as in Eqs.~(\ref{eq:dis}) and (\ref{eq:cov}). The above equation can be used to obtain the Wigner characteristic 
	function of a single mode coherent state: 
	\begin{equation}\label{chi_coh}
		\chi_\text{coh}(\Lambda)= \exp \left[-\frac{1}{4}(\tau ^2+\sigma ^2)-i (\tau  d_p-\sigma  d_x)\right],
	\end{equation}
	where $d_x$ and $d_p$ are related to the displacement vector as  $\bm{d}=(d_x,d_p)^T$. 
	Similarly, we can obtain the Wigner characteristic function of a single mode squeezed vacuum state with squeezing $r$ from Eq.~(\ref{wigc}):
	\begin{equation}\label{chi_sqv}
		\chi_\text{sqv}(\Lambda)=\exp \left[-\frac{1}{4}(\tau ^2 e^{2r}+\sigma ^2 e^{-2r}) \right].
	\end{equation}

	Consider our system of interest being subjected to a symplectic transformation $S$. The state of our system represented by the density operator $\hat{\rho}$ will then change to $\rho \rightarrow \,\mathcal{U}(S)\, \hat{\rho}
	\,\mathcal{U}(S)^{\dagger}$, where $\mathcal{U}(S)$ is the infinite dimensional unitary operator corresponding to the symplectic transformation $S$. 
	Under the symplectic transformation $S$, the first and second moments transform as $\bm{d}\rightarrow S \bm{d}$ and $\quad V\rightarrow SVS^T$, respectively. 
	The Wigner characterisic function changes from $\chi(\Lambda)$ to $\chi(S^{-1}\Lambda)$.

	\section{Matrices}\label{appdix}
	\begin{equation}\label{m1}
		M_1=\frac{-1}{4a_0}\left(
		\begin{array}{cccc}
			a_1 & 0 & a_2 & 0 \\
			0 & a_1 & 0 & -a_2 \\
			a_2 & 0 & a_3 & 0 \\
			0 & -a_2 & 0 & a_3 \\
		\end{array}
		\right),
	\end{equation}
	where 
	\begin{equation}
		\begin{aligned}
			a_0=&4 \kappa ^2 r_1^2 r_2^2+\gamma  \left(1-t_1^2 t_2^2\right)+\Gamma _1 \Gamma _2\\
			a_1=&4 \Gamma _1 \kappa ^2 r_2^2+\gamma  \left(1+t_1^2 t_2^2\right)+\Gamma _2 r_1^2\\
			a_2=&-16 \alpha  \beta  \kappa  t_1 t_2\\
			a_3=&4 \Gamma _2 \kappa ^2 r_1^2+\gamma  \left(1+t_1^2 t_2^2\right)+\Gamma _1 r_2^2\\
		\end{aligned}
	\end{equation}
	Here  $\Gamma _i=(1+T_i)$, $t_i=\sqrt{T_i}$, and $r_i=\sqrt{1-T_i}$ ($i=1,2$). 
	Further $\alpha=\sinh \, r$, $\beta=\cosh \, r$, and $\gamma=4\kappa(\alpha^2+\beta^2)$.
	\begin{equation}\label{m2}
		M_2=\frac{1}{a_0}\left(
		\begin{array}{cccc}
			b_1 & i b_1 & b_2 & -i b_2 \\
			b_3 & i b_1 & -b_2 & -i b_2 \\
			b_4 & -i b_4 & b_5 & i b_5 \\
			-b_4 & -i b_4 & b_6 & i b_5 \\
			b_7 & i b_7 & b_8 & -i b_8 \\
			-b_7 & i b_7 & -b_8 & -i b_8 \\
			b_9 & -i b_9 & b_{10} & i b_{10} \\
			-b_9 & -i b_9 & -b_{10} & i b_{10} \\
		\end{array}
		\right),
	\end{equation}
	where
	\begin{equation}
		\begin{aligned}
			b_1 = & r_1 \left(\gamma +\Gamma _2+4 \kappa ^2 r_2^2\right) \\
			b_2 = & -8 \alpha  \beta  \kappa  r_1 t_1 t_2 \\
			b_3  = & -r_1 \left(\gamma +\Gamma _2+4 \kappa ^2 r_2^2\right) \\
			b_4  = & -8 \alpha  \beta  \kappa  r_2 t_1 t_2 \\
			b_5  = & r_2 \left(\gamma +\Gamma _1+4 \kappa ^2 r_1^2\right) \\
			b_6  = & -r_2 \left(\gamma +\Gamma _1+4 \kappa ^2 r_1^2\right) \\
			b_7  = & r_1 t_1 \left(\Gamma _2+4 \kappa ^2 r_2^2-\gamma  t_2^2\right) \\
			b_8  = & 8 \alpha  \beta  \kappa  r_1 t_2 \\
			b_9  = & 8 \alpha  \beta  \kappa  r_2 t_1 \\
			b_{10}  = & r_2 t_2 \left(\Gamma _1+4 \kappa ^2 r_1^2-\gamma  t_1^2\right) \\
		\end{aligned}
	\end{equation}
	
	\begin{equation}\label{m3}
		M_3=\frac{1}{a_0}\left(
		\begin{array}{cccccccc}
			0 & c_1 & c_2 & 0 & 0 & c_3 & c_4 & 0 \\
			c_1 & 0 & 0 & c_2 & c_3 & 0 & 0 & c_4 \\
			c_2 & 0 & 0 & c_5 & c_6 & 0 & 0 & c_7 \\
			0 & c_2 & c_5 & 0 & 0 & c_6 & c_7 & 0 \\
			0 & c_3 & c_6 & 0 & 0 & c_8 & c_9 & 0 \\
			c_3 & 0 & 0 & c_6 & c_8 & 0 & 0 & c_9 \\
			c_4 & 0 & 0 & c_7 & c_9 & 0 & 0 & c_{10} \\
			0 & c_4 & c_7 & 0 & 0 & c_9 & c_{10} & 0 \\
		\end{array}
		\right)
	\end{equation}
	where
	\begin{equation}
		\begin{aligned}
			c_1  = & r_1^2 \left(\gamma +\Gamma _2+4 \kappa ^2 r_2^2\right) \\
			c_2  = & 8 \alpha  \beta  \kappa  r_1 r_2 t_1 t_2 \\
			c_3  = & t_1 \left(2 \Gamma _2+\gamma  r_2^2\right) \\
			c_4  = & -8 \alpha  \beta  \kappa  r_1 r_2 t_1 \\
			c_5  = & r_2^2 \left(\gamma +\Gamma _1+4 \kappa ^2 r_1^2\right) \\
			c_6  = & -8 \alpha  \beta  \kappa  r_1 r_2 t_2 \\
			c_7  = & t_2 \left(2 \Gamma _1+\gamma  r_1^2\right) \\
			c_8  = & r_1^2 \left(-\Gamma _2+4 \kappa ^2 r_2^2+\gamma  t_2^2\right) \\
			c_9  = & 8 \alpha  \beta  \kappa  r_1 r_2 \\
			c_{10}  = & r_2^2 \left(-\Gamma _1+4 \kappa ^2 r_1^2+\gamma  t_1^2\right) \\
		\end{aligned}
	\end{equation}

 %%%%%%%%%%%%%%%%%%%%%%%%%%%%%%%%%%%%%%%%%%%%%
 \newpage
		
 %\bibliography{references}
%apsrev4-2.bst 2019-01-14 (MD) hand-edited version of apsrev4-1.bst
%Control: key (0)
%Control: author (8) initials jnrlst
%Control: editor formatted (1) identically to author
%Control: production of article title (0) allowed
%Control: page (0) single
%Control: year (1) truncated
%Control: production of eprint (0) enabled
%

%%%%%%%%%%%%%%%%%%%%%%%%%%%%%%%%%%%%%%%%%%%%%%%%%	

\end{document}